\documentclass[amsmath,amssymb,aps,prl,twocolumn,groupedaddress,10pt]{revtex4-2}
\usepackage{bm}
\usepackage{dcolumn}
\usepackage[pdftex]{graphicx}
\usepackage[pdftex,pdfborder={0 0 0}]{hyperref}
\usepackage{lineno}
\usepackage{moreverb}
\usepackage{multirow}
\usepackage{xcolor}
\hypersetup{
    colorlinks=true,
    citecolor=blue,
    linkcolor=blue,
    urlcolor=blue,
}

\begin{document}


\title{Observation of Coulomb-Assisted Nuclear Bound State of $\Xi^-$--$^{14}$N System}


\author{S.~H.~Hayakawa$^1$}
\email{hayashu@post.j-parc.jp}
\author{K.~Agari$^2$}
\author{J.~K.~Ahn$^3$}
\author{T.~Akaishi$^4$}
\author{Y.~Akazawa$^2$}
\author{S.~Ashikaga$^{1,5}$}
\author{B.~Bassalleck$^6$}
\author{S.~Bleser$^7$}
\author{H.~Ekawa$^8$}
\author{Y.~Endo$^9$}
\author{Y.~Fujikawa$^5$}
\author{N.~Fujioka$^{10}$}
\author{M.~Fujita$^{1}$}
\author{R.~Goto$^9$}
\author{Y.~Han$^{11}$}
\author{S.~Hasegawa$^1$}
\author{T.~Hashimoto$^1$}
\author{T.~Hayakawa$^4$}
\author{E.~Hayata$^5$}
\author{K.~Hicks$^{12}$}
\author{E.~Hirose$^2$}
\author{M.~Hirose$^5$}
\author{R.~Honda$^{2}$}
\author{K.~Hoshino$^9$}
\author{S.~Hoshino$^4$}
\author{K.~Hosomi$^1$}
\author{S.~H.~Hwang$^{13}$}
\author{Y.~Ichikawa$^1$}
\author{M.~Ichikawa$^{5,14}$}
\author{K.~Imai$^1$}
\author{K.~Inaba$^5$}
\author{Y.~Ishikawa$^{10}$}
\author{H.~Ito$^9$}
\author{K.~Ito$^{15}$}
\author{W.~S.~Jung$^3$}
\author{S.~Kanatsuki$^5$}
\author{H.~Kanauchi$^{10}$}
\author{A.~Kasagi$^{8,16}$}
\author{T.~Kawai$^{17}$}
\author{M.~H.~Kim$^3$}
\author{S.~H.~Kim$^3$}
\author{S.~Kinbara$^{16}$}
\author{R.~Kiuchi$^{18}$}
\author{H.~Kobayashi$^9$}
\author{K.~Kobayashi$^4$}
\author{T.~Koike$^{10}$}
\author{A.~Koshikawa$^5$}
\author{J.~Y.~Lee$^{19}$}
\author{T.~L.~Ma$^{20}$}
\author{S.~Y.~Matsumoto$^{5,14}$}
\author{M.~Minakawa$^2$}
\author{K.~Miwa$^{10}$}
\author{A.~T.~Moe$^{21}$}
\author{T.~J.~Moon$^{19}$}
\author{M.~Moritsu$^2$}
\author{Y.~Nagase$^9$}
\author{Y.~Nakada$^4$}
\author{M.~Nakagawa$^8$}
\author{D.~Nakashima$^9$}
\author{K.~Nakazawa$^{9,16}$}
\author{T.~Nanamura$^{1,5}$}
\author{M.~Naruki$^{1,5}$}
\author{A.~N.~L.~Nyaw$^{16}$}
\author{Y.~Ogura$^{10}$}
\author{M.~Ohashi$^9$}
\author{K.~Oue$^4$}
\author{S.~Ozawa$^{10}$}
\author{J.~Pochodzalla$^{7,22}$}
\author{S.~Y.~Ryu$^{23}$}
\author{H.~Sako$^1$}
\author{S.~Sato$^1$}
\author{Y.~Sato$^2$}
\author{F.~Schupp$^7$}
\author{K.~Shirotori$^{23}$}
\author{M.~M.~Soe$^{24}$}
\author{M.~K.~Soe$^{16}$}
\author{J.~Y.~Sohn$^{25}$}
\author{H.~Sugimura$^{26}$}
\author{K.~N.~Suzuki$^5$}
\author{H.~Takahashi$^2$}
\author{T.~Takahashi$^2$}
\author{T.~Takeda$^5$}
\author{H.~Tamura$^{1,10}$}
\author{K.~Tanida$^1$}
\author{A.~M.~M.~Theint$^{16}$}
\author{K.~T.~Tint$^9$}
\author{Y.~Toyama$^{10}$}
\author{M.~Ukai$^{2,10}$}
\author{E.~Umezaki$^5$}
\author{T.~Watabe$^{15}$}
\author{K.~Watanabe$^5$}
\author{T.~O.~Yamamoto$^1$}
\author{S.~B.~Yang$^3$}
\author{C.~S.~Yoon$^{25}$}
\author{J.~Yoshida$^{8,10}$}
\author{M.~Yoshimoto$^9$}
\author{D.~H.~Zhang$^{20}$}
\author{Z.~Zhang$^{20}$\vspace{10pt}}

\affiliation{$^1$Advanced Science Research Center, Japan Atomic Energy Agency, Tokai 319-1195, Japan}
\affiliation{$^2$Institute of Particle and Nuclear Studies, High Energy Accelerator Research Organization (KEK), Tsukuba 305-0801, Japan}
\affiliation{$^3$Department of Physics, Korea University, Seoul 02841, Korea}
\affiliation{$^4$Department of Physics, Osaka University, Toyonaka 560-0043, Japan}
\affiliation{$^5$Department of Physics, Kyoto University, Kyoto 606-8502, Japan}
\affiliation{$^6$Department of Physics and Astronomy, University of New Mexico, Albuquerque, New Mexico 87131, USA}
\affiliation{$^7$Helmholtz Institute Mainz, 55099 Mainz, Germany}
\affiliation{$^8$High Energy Nuclear Physics Laboratory, RIKEN, Wako 351-0198, Japan}
\affiliation{$^9$Faculty of Education, Gifu University, Gifu 501-1193, Japan}
\affiliation{$^{10}$Department of Physics, Tohoku University, Sendai 980-8578, Japan}
\affiliation{$^{11}$Institute of Nuclear Energy Safety Technology, Hefei Institutes of Physical Science, Chinese Academy of Sciences, Hefei 230031, China}
\affiliation{$^{12}$Department of Physics \& Astronomy, Ohio University, Athens, Ohio 45701, USA}
\affiliation{$^{13}$Korea Research Institute of Standards and Science, Daejeon 34113, Korea}

\affiliation{$^{14}$Meson Science Laboratory, RIKEN, Wako 351-0198, Japan}

\affiliation{$^{15}$Department of Physics, Nagoya University, Nagoya 464-8601, Japan}

\affiliation{$^{16}$Graduate School of Engineering, Gifu University, Gifu 501-1193, Japan}

\affiliation{$^{17}$Center for Advanced Photonics, RIKEN, Wako 351-0198, Japan}
\affiliation{$^{18}$Institute of High Energy Physics, Beijing 100049, China}
\affiliation{$^{19}$Department of Physics, Seoul National University, Seoul 08826, Korea}
\affiliation{$^{20}$Institute of Modern Physics, Shanxi Normal University, Linfen 041004, China}
\affiliation{$^{21}$Department of Physics, Lashio University, Lashio 06301, Myanmar}
\affiliation{$^{22}$Institut fur Kernphysik, Johannes Gutenberg-Universitat, 55099 Mainz, Germany}
\affiliation{$^{23}$Research Center for Nuclear Physics, Osaka University, Osaka 567-0047, Japan}
\affiliation{$^{24}$Department of Physics, University of Yangon, Yangon 11041, Myanmar}
\affiliation{$^{25}$Research Institute of Natural Science, Gyeongsang National University, Jinju 52828, Korea}
\affiliation{$^{26}$Accelerator Laboratory, High Energy Accelerator Research Organization (KEK), Tsukuba 305-0801, Japan}

\collaboration{J-PARC E07 Collaboration}
\noaffiliation

\date{\today}

\begin{abstract}
In an emulsion-counter hybrid experiment performed at J-PARC,
a $\Xi^-$ absorption event was observed which decayed into twin
single-$\Lambda$ hypernuclei.
Kinematic calculations enabled a unique identification of the reaction process as
$\Xi^{-} + ^{14}$N$\ \rightarrow\ ^{10}_\Lambda$Be + $^5_\Lambda$He.
For the binding energy of the $\Xi^{-}$ hyperon in
the $\Xi^-$--$^{14}$N system a value of $1.27 \pm 0.21$ MeV was deduced.
The energy level of $\Xi^-$ is likely a nuclear $1p$ state which
indicates a weak ${\Xi}N$--$\Lambda\Lambda$ coupling.
\end{abstract}


\maketitle
\thispagestyle{plain}


Strange nuclear systems cover a broad spectrum of phenomena. They range from two-body
baryon-baryon interactions, over complex nuclei containing strange baryons, strangeness in hot nuclear
systems created in heavy ion reactions, up to perhaps the inner core of neutron stars.
In contrast to a relatively large amount of information on $S = -1$
$\Lambda$-hypernuclei, experimental data regarding $S = -2$ systems are still scarce.
Double-$\Lambda$ hypernuclei have represented the preferred method to study
the $\Lambda$--$\Lambda$ interaction, and a pioneering
binding energy determination of $_{\Lambda\Lambda}^6$He
revealed the $\Lambda\Lambda$ interaction to be weakly attractive \cite{nagara2, e373}.
Only recently, the $\Xi^-$--$p$ interaction was studied by ALICE \cite{alice, alice2020}.
From the two-body correlations, the presence of a strongly attractive interaction was inferred.

In line with single $\Lambda$ hypernuclei, the study of $\Xi$ hypernuclei can provides
meaningful information on the $\Xi{N}$ interaction.
First $(K^-, K^+)$ missing-mass spectroscopy studies
were performed by the KEK E224 and BNL E885 collaborations.
In both experiments, insufficient energy resolution prevented the observation of a peak in the bound state region
\cite{e224, e885}.
Assuming a Woods-Saxon type potential, the BNL E885 experiment estimated the potential depth of the $\Xi^-$
to be about 14 MeV which suggests a binding energy around 4.5 MeV.
While an initial J-PARC experiment \cite{e05} measured the missing mass spectrum of
the $^{12}$C$(K^-, K^+)$ reaction,
a new experiment with a much improved energy resolution of better than 2 MeV FWHM is now planned at J-PARC \cite{e70}.

Several emulsion experiments reported the possibility of an attractive
$\Xi$-nucleus interaction. A remarkable event named ``KISO'' was found
by the KEK E373 experiment \cite{kiso}.
The decay mode of that event was uniquely identified to be
$\Xi^{-}$ + $^{14}$N $\rightarrow$ $^{10}_\Lambda$Be + $^5_\Lambda$He.
For the binding energy of the $\Xi^-$ hyperon, $B_{\Xi^-}$, a value of $3.87 \pm 0.21$\,MeV
or $1.03 \pm 0.18$\,MeV was deduced, depending whether the $^{10}_\Lambda$Be daughter nucleus is produced in
the ground state or the excited state, respectively \cite{hiyama-nakazawa}.
In either scenario, the bound state of the $\Xi^-$--$^{14}$N system is expected
to be deeper than the atomic $3D$ orbit.

The $\Xi{N}$ interaction can also be extracted
by measuring the energy shift and width of x rays from $\Xi$ atoms.
Two experiments involving $\Xi$-atomic x ray measurements
using Ge detectors have been proposed at J-PARC \cite{e03, e07},
E07 being the one described in this paper.

A theoretical calculation of the binding energy of
the $\Xi^-$--$^{14}$N system was presented by Yamaguchi {\it et al.}
using the ${\Xi}N$ one-boson-exchange
potential called the Ehime potential \cite{ehime-2}.
In this model, the coupling constants were adjusted
to reproduce the experimental result of the $\Xi^-$--$^{12}$C
bound states with $B_{\Xi^-} \sim$ 0.6 MeV observed
in the KEK E176 experiment \cite{e176-rev, hiyama-nakazawa}.
The calculation also predicted for the $\Xi^-$--$^{11}$B system
a ground state binding energy, which is in agreement with the excitation energy spectrum
in the BNL E885 experiment. More recently, T. T. Sun {\it et al.} performed
a theoretical calculation with the relativistic-mean-field
and Skyrme-Hartree-Fock models \cite{ttsun}.
The preferred interpretation of the KISO event was an observation of
an excited state of the $^{10}_\Lambda$Be.
When the ${\Xi}N$ interaction was adjusted to reproduce
the binding energy for the KISO event assuming the $1p$ state
for an excited $_\Lambda^{10}$Be,
the predicted $\Xi^-$ removal energy of $^{15}_\Xi$C in the $1s$ state was
7.2--9.4 MeV.
Very recent Lattice QCD calculations with almost physical quark
masses ($m_\pi = 146$ MeV), provided the ${\Xi}N$ interaction potentials for various $S = -2$
channels \cite{lattice}. These lattice results indicated that the coupling
between $\Lambda\Lambda$ and ${\Xi}N$ states is weak.

J-PARC E07 is an emulsion-counter hybrid experiment
aiming to identify the decay modes of about 10 events of
$S = -2$ hypernuclei \cite{e07}.
The experiment was carried out using
a 1.81 GeV/$c$ $K^-$ beam at the K1.8 beam line of the
Hadron Experimental Facility at J-PARC \cite{hadron, k18bs}.
The $\Xi^-$ hyperons produced in the quasifree
``$p$''$(K^-, K^+)\Xi^-$ reaction in a diamond target
of 9.87 g/cm$^2$ thickness were injected into an emulsion module
located downstream of the target.
The emulsion module consisted of two 380-$\mu$m-thick sheets and
eleven 1-mm-thick sheets with $34.5 \times 35.0$ cm$^2$ area.
The incident $\Xi^-$ hyperons were eventually slowed down and captured
at rest in the atomic orbit of a nucleus in the emulsion material.
$\Xi$ hypernuclei or double-$\Lambda$ hypernuclei are generated
at the capture point with some probability \cite{ammt},
and the decay tracks of charged particles
are recorded in the emulsion module.
In total, 118 emulsion modules were exposed to $1.13 \times 10^{11}$
$K^-$ particles. About 100 events of $S = -2$ hypernuclei were expected
to be produced as a result of the $10^4$ $\Xi^-$ hyperons stopped in the emulsion.
More details on the experimental setup can be found in Ref.\,\cite{mino}.
The $\Xi$-atomic x rays were also measured by using germanium detectors.
Details are presented in Ref.\,\cite{fujita}.

\begin{figure}[ht]
\centering
\includegraphics[clip,width=0.4\textwidth]{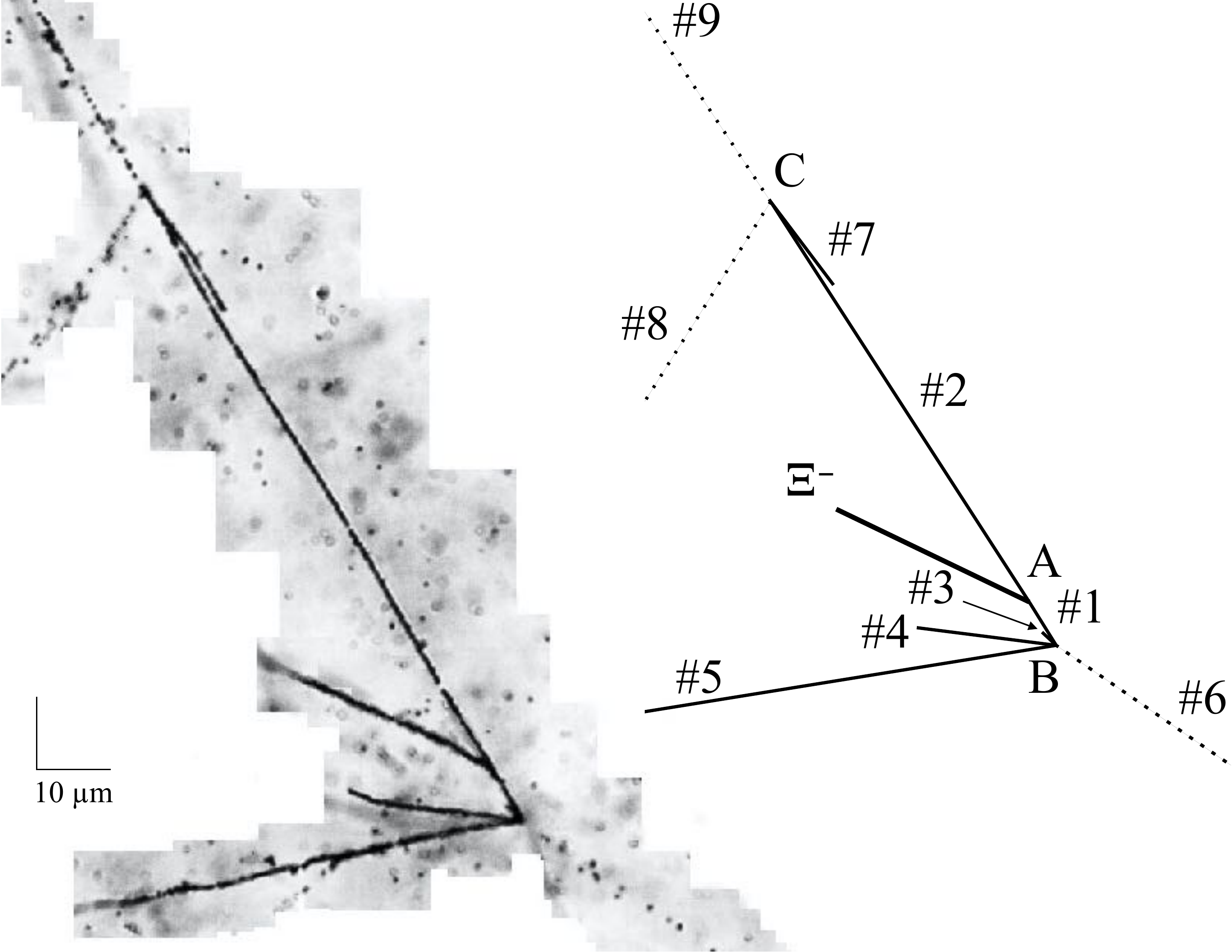}
\caption{\label{fig:fig1}
  Superimposed image and schematic drawing of the IBUKI event.}
\end{figure}

A remarkable event forming a twin-$\Lambda$ hypernuclear topology
was found in the tenth sheet of module \#047.
Figure \ref{fig:fig1} shows a superimposed image and a schematic drawing
of the event. We named the event ``IBUKI''
\footnote{IBUKI is the name of a mountain on the east-west boundary of the
Japanese main island.}.
The $\Xi^-$ traced in sequence from upstream was found to have stopped
and decayed at vertex $A$, from which the two charged particles
of tracks \#1 and \#2 were emitted.
Track \#1 decayed into four charged particles, tracks \#3--6 at vertex $B$.
Track \#2 showed a decay into three charged particles,
tracks \#7--9 at vertex $C$.
All nine tracks were manually followed and their stopping points were found
inside of the emulsion module.
No charged daughter particles were found at the ends of tracks \#5, \#6, and \#9.
Track \#8 accompanied by particle emission at the stopping point
indicates a negative particle.
Additional support for this comes from the observed Auger emission.
Table \ref{table:range-summary} summarizes the measured values
of ranges and emission angles of the nine tracks.
\begin{table*}[ht]
\caption{\label{table:range-summary}
  Table of the measured ranges and angles of the nine tracks.}
\begin{ruledtabular}
  \begin{tabular}{ccD{+}{\,\pm\,}{-1}D{+}{\,\pm\,}{-1}D{+}{\,\pm\,}{-1}c}
    Vertex & Track & \multicolumn{1}{c}{Range [$\mu$m]} &
    \multicolumn{1}{c}{$\theta$ [deg]} & \multicolumn{1}{c}{$\phi$ [deg]} &
    Comment \\ \hline
    $A$ & \#1 &     8.2 +  0.4 &  90.6 + 1.9 & 301.9 + 2.1 & $\Lambda$
    hypernucleus \\
      & \#2 &    88.5 +  0.3 &  93.5 + 2.0 & 122.9 + 1.5 & $\Lambda$
    hypernucleus \\
    $B$ & \#3 &     3.6 +  1.1 & 124.6 + 7.4 & 137.0 + 6.4 \\
      & \#4 &    25.9 +  0.3 & 109.8 + 2.3 & 172.8 + 1.5 \\
      & \#5 &   961.1 +  4.6 &  49.6 + 2.0 & 189.2 + 2.0 \\
      & \#6 & 20509   + 34   & 111.6 + 2.7 & 325.5 + 2.2 \\
    $C$ & \#7 &    19.2 +  0.1 &  86.8 + 1.6 & 307.0 + 1.2 \\
      & \#8 &  2159   + 19   &  30.6 + 1.5 & 237.8 + 2.8 &
    $\pi^-$ with $\sigma$ stop \\
      & \#9 &  2179.0 +  3.6 & 106.7 + 1.6 & 123.5 + 1.2 \\
  \end{tabular}
\end{ruledtabular}
\end{table*}
The angles are expressed by a zenith angle ($\theta$) and an azimuthal
angle ($\phi$) with respect to the axis perpendicular to the emulsion
sheet.
The ranges in the base film were converted to those in the emulsion layer.

Since the energy calibration and the range correction were
necessary for the kinematic analysis,
the density of the emulsion was measured
using 132 $\alpha$ tracks with a monochromatic energy of 8.785 MeV
from the decay of $^{212}$Po.
The mean range and the emulsion density were measured to be
$50.25 \pm 0.11$ $\mu$m and $3.544 \pm 0.012$ g/cm$^3$, respectively.

For each vertex point,
all possible decay modes were kinematically examined considering
both mesonic and non-mesonic decays.
Among various nuclear species in the emulsion,
C, N, and O were taken into account as nuclei to capture the $\Xi^-$ hyperon.
Emission of neutral particles was also considered.
As for the mass of the $\Lambda$ hypernuclei,
the values obtained from the experimental data were utilized
\cite{hashimoto-tamura, lmass-1, lmass-2, lmass-3, lmass-4,
  lmass-5, lmass-6, lmass-8}.
Mass values of some possible $\Lambda$ hypernuclei were also
taken into account via a calculation with linear interpolation and
extrapolation of $B_\Lambda$, where a typical error in the fitting of
the masses of 0.5 MeV/$c^2$ was uniformly assumed.

\begin{figure}[ht]
\centering
\includegraphics[clip,width=0.42\textwidth]{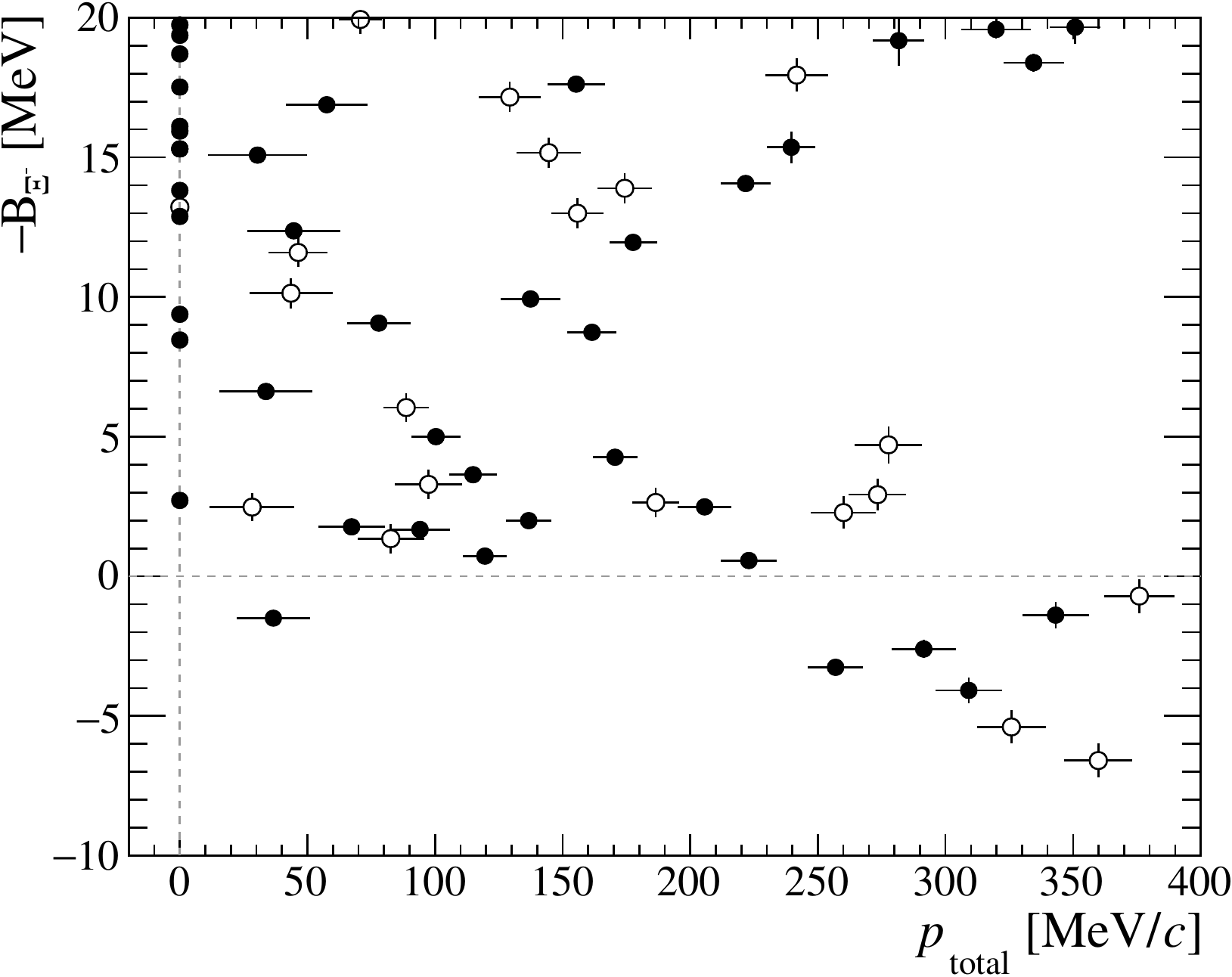}
\caption{\label{fig:fig2}
  Correlation plot of the binding energy and the magnitude of
  the total momentum at vertex $A$.
  Black dots and open circles indicate the decay modes including known
  and possible $\Lambda$ hypernuclei, respectively.
}
\end{figure}

Figure \ref{fig:fig2} shows $B_{\Xi^-}$,
and the magnitude of the total momentum, $p_{\rm total}$,
for all possible decay modes for vertex $A$.
The black dots and open circles indicate the decay modes including known and
possible $\Lambda$ hypernuclei, respectively.
The $p_{\rm total}$ value was required to be zero within 3$\sigma$ tolerance.
Since the stopped $\Xi^-$ cascades down atomic orbitals before being absorbed
by the nucleus, the binding energy will be positive, at least within
the experimental error.
The value of $B_{\Xi^-}$ was required to be more than zero
with 3$\sigma$ tolerance, even if multiple neutral particles are emitted.
Finally, only one decay mode of
\begin{equation*}\label{eqn:vertexA-mode}
  \Xi^- + {}^{14}\text{N}\ \rightarrow\ {}^{10}_\Lambda\text{Be}(\#1)
  + {}^{5}_\Lambda\text{He}(\#2)
\end{equation*}
was accepted around the position of $p_{\rm total} \sim 40$ MeV$/c$ and
$B_{\Xi^-} \sim 1 $ MeV in Figure \ref{fig:fig2}.

At vertex $B$, the kinematic calculation was also performed using tracks \#3--6.
358 possible decay modes remained within 3$\sigma$ tolerance of the energy and
the momentum conservation law independent of the result of vertex $A$.
Among them, 13 decay modes for the case of track \#1 being
${}_\Lambda^{10}$Be are listed in Table \ref{table:vertexB}.

\begin{table}[htbp]
  \centering
  \caption[Possible decay modes at vertex $B$
    for the case of track \#1 being ${}^{10}_\Lambda$Be]
          {Possible decay modes at vertex $B$
            for the case of track \#1 being ${}^{10}_\Lambda$Be.}
\begin{ruledtabular}
  \begin{tabular}{ccccccc}
    \#3 & \#4 & \#5 & \#6 & 
     & $Q$ [MeV] & $B_\Lambda$ [MeV] \\
    \hline
    $p$ & $d$ & $t$ & $p$ & 3$n$ & 120.06 & $<$ 11.81 \\
    $p$ & $t$ & $d$ & $p$ & 3$n$ & 120.06 & $<$ 16.36 \\
    $p$ & $t$ & $t$ & $p$ & 2$n$ & 126.32 & $<$ 12.83 \\
    $d$ & $p$ & $t$ & $p$ & 3$n$ & 120.06 & $<$ 12.71 \\
    $d$ & $d$ & $d$ & $p$ & 3$n$ & 116.03 & $<$ 12.89 \\
    $d$ & $d$ & $t$ & $p$ & 2$n$ & 122.28 & $<$ 10.04 \\
    $d$ & $t$ & $p$ & $p$ & 3$n$ & 120.06 & $<$ 20.07 \\
    $d$ & $t$ & $d$ & $p$ & 2$n$ & 122.28 & $<$ 15.27 \\
    $t$ & $p$ & $d$ & $p$ & 3$n$ & 120.06 & $<$ 17.31 \\
    $t$ & $p$ & $t$ & $p$ & 2$n$ & 126.32 & $<$ 15.09 \\
    $t$ & $d$ & $p$ & $p$ & 3$n$ & 120.06 & $<$ 20.15 \\
    $t$ & $d$ & $d$ & $p$ & 2$n$ & 122.29 & $<$ 15.88 \\
    $t$ & $t$ & $p$ & $p$ & 2$n$ & 126.32 & $<$ 22.48 \\
  \end{tabular}
  \label{table:vertexB}
\end{ruledtabular}
\end{table}

Since the length of track \#3 was very short,
the topology without track \#3 was also tested.
The coplanarity of three tracks is defined as
$\left( \hat{r}_1 \times \hat{r}_2 \right) \hat{r}_3$,
where $\hat{r}_i$ represents the unit vector of $i$th track direction.
The coplanarity of tracks \#4--6 was obtained to be $- 0.50 \pm 0.03$.
This makes some neutron(s) emission likely.
The accepted decay modes for the case of track \#1 being ${}_\Lambda^{10}$Be
are listed in Table \ref{table:vertexBwo3}.
From 48 candidates,
the listed three decay modes were consistent with the result at vertex $A$.
Since no track from $\beta$ decay of $^6$He to $^6$Li $+$ $e^-$ $+$ $\bar{\nu}$
is visible at the end of track \#4, the case of track \#4 being $^6$He was
rejected.

\begin{table}[htbp]
  \centering
  \caption[Possible decay modes at vertex $B$ without track \#3
    for the case of track \#1 being ${}^{10}_\Lambda$Be]
          {Possible decay modes at vertex $B$ without track \#3
            for the case of track \#1 being ${}^{10}_\Lambda$Be.}
\begin{ruledtabular}
  \begin{tabular}{ccccccc}
    \#4 & \#5 & \#6 & 
     & $Q$ [MeV] & $B_\Lambda$ [MeV] \\
    \hline
    $^3$He & $t$ & $p$ & 3$n$ & 125.55 & $<$ 12.13 \\
    $^4$He & $d$ & $p$ & 3$n$ & 139.87 & $<$ 31.13 \\
    $^4$He & $t$ & $p$ & 2$n$ & 146.13 & $<$ 24.69 \\
    $^6$He & $p$ & $p$ & 2$n$ & 138.63 & $<$ 27.68 & rejected \\
  \end{tabular}
  \label{table:vertexBwo3}
\end{ruledtabular}
\end{table}

For vertex $C$, consistency with the conservation laws was checked using
tracks \#7--9.
In total, 263 candidates were accepted within 3$\sigma$ tolerance of energy
and momentum conservation.
Almost all were nonmesonic decay modes with multiple neutral particles.
Among the 263 candidates,
the decay mode for the case of track \#2 being ${}_\Lambda^{5}$He was
the only reasonable one, obtained as the following decay mode,
\begin{equation*}\label{eqn:vertexC-mode}
  ^5_\Lambda\text{He}\ \rightarrow\ {}^{4}\text{He}(\#7)
  + \pi^{-}(\#8) + p(\#9).
\end{equation*}
This decay mode is consistent with a charged-particle emission at
the end point of track \#8.
Since the value of the coplanarity of tracks\#7--9 was $0.06 \pm 0.03$
at vertex $C$,
no neutral particle was likely emitted at vertex $C$.

From the above discussions, the reaction process of the IBUKI event
was obtained as follows,
\begin{eqnarray*}\label{eqn:ibuki-mode}
  \Xi^- + {}^{14}\text{N}&\rightarrow&{}^{10}_\Lambda\text{Be}(\#1)
  + {}^{5}_\Lambda\text{He}(\#2), \\
  ^{10}_\Lambda\text{Be}&\rightarrow& \text{(3 or 4 nuclei, \#3--6)}
  + \text{(2 or 3 } n),\\
  ^5_\Lambda\text{He}&\rightarrow&{}^{4}\text{He}(\#7)
  + \pi^{-}(\#8) + p(\#9).
\end{eqnarray*}
Thus, the formation process of twin $\Lambda$ hypernuclei was uniquely
identified as $^{10}_\Lambda\text{Be}$ and ${}^{5}_\Lambda\text{He}$.
Since the final state was a two-body system at vertex $A$,
the momenta of the twin $\Lambda$ hypernuclei
should balance if the initial state is at rest.
The range from vertexes $B$ to $C$ was measured to be $97.1 \pm 0.4$ $\mu$m.
To minimize the measurement errors,
a kinematic fitting was applied at vertex $A$ \cite{kinematic-fitting}.
The fitting result is summarized in Table \ref{table:kinema-fitting-A}.
From the constraint of the total length between vertex $B$ and $C$,
the ranges of track \#1 and \#2 were calculated to be
$9.4 \pm 0.2$ $\mu$m and $87.7 \mp 0.2$ $\mu$m, respectively,
so that their momenta correspond to each other.
Straggling was taken into account in this fitting.
The angles represent twin $\Lambda$ hypernuclei being emitted back to back.
The $\chi^2$ value was obtained to be 2.4 with 3 degrees of freedom.
Considering the effects of straggling and the error in the emulsion density,
the error of $B_{\Xi^-}$ was estimated to be $0.08$ MeV.
The mass errors of $\Xi^-$ hyperon (1321.71 $\pm$ 0.07 MeV/$c^2$\cite{ximass}),
${}^{10}_\Lambda$Be (9499.88 $\pm$ 0.13 MeV/$c^2$\cite{gogami-be10}), and
${}^{5}_\Lambda$He (4839.94 $\pm$ 0.02 MeV/$c^2$\cite{lmass-1}) were also
taken into account.
The mass of ${}^{10}_\Lambda$Be was evaluated as the average of
the $1^-$ and $2^-$ states in the ground-state doublet.
The energy spacing between these two states is expected to be less than
0.1 MeV.
Therefore,
the binding energy of the $\Xi^-$ hyperon in the $\Xi^-$--${}^{14}$N system,
$B_{\Xi^-}$ was obtained to be $1.27 \pm 0.21$ MeV, where
the error includes the spin-doublet uncertainty of $^{10}_\Lambda$Be.
Kinematic fitting was also applied at vertex $C$.
The result is listed in Table \ref{table:kinema-fitting-C}.
The $\chi^2$ value was obtained to be 6.6 with 4 degrees of freedom.
The binding energy of the $\Lambda$ hyperon in $^{5}_\Lambda$He,
$B_\Lambda$, was obtained to be 2.77 $\pm$ 0.23 MeV, which agrees well
with the world average \cite{lmass-1}.

\begin{table}[htbp]
  \centering
  \caption[Ranges and emission angles for vertex $A$ with the kinematic fitting]
          {Ranges and emission angles for vertex $A$ with the kinematic fitting.
           The value of $\chi^2/\text{ndf}$ is $2.4/3$.}
\begin{ruledtabular}
  \begin{tabular}{ccD{+}{\,\pm\,}{-1}D{+}{\,\pm\,}{-1}D{+}{\,\pm\,}{-1}c}
    \multicolumn{1}{c}{Vertex} & \multicolumn{1}{c}{Track} &
    \multicolumn{1}{c}{Range [$\mu$m]} & \multicolumn{1}{c}{$\theta$ [deg]} &
    \multicolumn{1}{c}{$\phi$ [deg]} \\
    \hline
    $A$ & \#1 & 9.4 + 0.1 & 88.7 + 1.4 & 302.5 + 1.2 \\
    & \#2 & 87.7 + 0.7 & 91.4 + 1.4 & 122.5 + 1.2 \\
  \end{tabular}
  \label{table:kinema-fitting-A}
\end{ruledtabular}
\end{table}

\begin{table}[htbp]
  \centering
  \caption[Ranges and emission angles for vertex $C$ with the kinematic fitting]
          {Ranges and emission angles for vertex $C$ with the kinematic fitting.
            The value of $\chi^2/\text{ndf}$ is $6.6/4$.}
\begin{ruledtabular}
  \begin{tabular}{ccD{+}{\,\pm\,}{-1}D{+}{\,\pm\,}{-1}D{+}{\,\pm\,}{-1}c}
    \multicolumn{1}{c}{Vertex} & \multicolumn{1}{c}{Track} &
    \multicolumn{1}{c}{Range [$\mu$m]} & \multicolumn{1}{c}{$\theta$ [deg]} &
    \multicolumn{1}{c}{$\phi$ [deg]} \\
    \hline
    $C$ & \#7 & 19.1 + 0.2 & 87.0 + 1.0 & 308.9 + 0.9 \\
    & \#8 & 2146 + 71 & 29.5 + 1.5 & 236.3 + 2.2 \\
    & \#9 & 2194 + 28 & 105.7 + 1.0 & 121.6 + 0.8 \\
  \end{tabular}
  \label{table:kinema-fitting-C}
\end{ruledtabular}
\end{table}

The $B_{\Xi^-}$ value of $1.27 \pm 0.21$ MeV represents
a bound $\Xi^-$--${}^{14}$N system.
However, the case of $^{10}_\Lambda$Be
being produced in an excited state must be considered.
In a missing mass experiment at JLab the energy spectrum of $^{10}_\Lambda$Be
was measured \cite{gogami-be10}.
A low-lying excited state at 2.78 $\pm$ 0.11 MeV was observed.
In case of the KISO event, the production of the $^{10}_\Lambda$Be in
that excited state could not be excluded so that the $\Xi^-$ binding energy
was not uniquely determined (see Table \ref{table:bxi-past}).
In case of the IBUKI event, the small $\Xi^-$ binding energy of 1.27 MeV
excludes a production of $^{10}_\Lambda$Be in the 2.78 MeV state.
Instead, the $^{10}_\Lambda$Be is produced in one of the levels of
the ground-state doublet.
Thus, the reaction process of the IBUKI event was determined to be
a bound state of $\Xi^-$--${}^{14}$N decaying into ground states
of both $^{10}_\Lambda$Be and $^{5}_\Lambda$He,
having a $B_{\Xi^-}$ value of $1.27 \pm 0.21$ MeV.
This is the first observation of a twin-$\Lambda$ hypernuclei event
in which the binding energy was precisely determined.

\begin{table}[htbp]
  \centering
  \caption[Summary of the binding energy of the $\Xi^-$ hyperon
    measured in the past and present experiments]
          {Summary of the binding energy of the $\Xi^-$ hyperon
    measured in the past and present experiments.}
\begin{ruledtabular}
  \begin{tabular}{cclllD{+}{\,\pm\,}{-1}c}
    \multicolumn{1}{c}{Event} & \multicolumn{1}{c}{Target} &
    \multicolumn{3}{c}{Decay mode} &
    \multicolumn{1}{c}{$B_{\Xi^-}$ [MeV]} &\\
    \hline
    KISO \cite{hiyama-nakazawa, kiso} & ${}^{14}$N &
    $^{10}_\Lambda$Be & $^{5}_\Lambda$He & & 3.87 + 0.21
    &  \\
    & ${}^{14}$N & $^{10}_\Lambda$Be* & $^{5}_\Lambda$He & & 1.03 + 0.18 \\
    IBUKI (present data) & ${}^{14}$N & $^{10}_\Lambda$Be & $^{5}_\Lambda$He & & 1.27 + 0.21 \\
  \end{tabular}
  \label{table:bxi-past}
\end{ruledtabular}
\end{table}

In the calculation of Yamaguchi {\it et al.}
using the Ehime potential \cite{ehime-2},
the $B_{\Xi^-}$ values for the $\Xi^-$--$^{14}$N system
are 5.93 and 1.14 MeV in the nuclear $1s$ (atomic $1S$)
and nuclear $1p$ (atomic $2P$) states, respectively.
The bound states of both the IBUKI and the KISO events are consistent with
the calculation for the $1p$ state.
In the nuclear $1p$ state both Coulomb and nuclear forces are at work,
resulting in a binding energy of 0.39 and 0.75 MeV, respectively,
according to this calculation.
Thus the result is a Coulomb-assisted nuclear bound state.
The calculated $B_{\Xi^-}$ of $^{15}_\Xi$C and $^{12}_\Xi$Be
by T. T. Sun {\it et al.}
\cite{ttsun} are also consistent with the experimental data.

From the above considerations,
in order to satisfy the experimental results of KEK E176, BNL E885,
the KISO event, and the IBUKI event,
the interpretation of the KISO event likely results
in $B_{\Xi^-} = 1.03 \pm 0.18$ MeV.
In that case,
the energy level of $\Xi^-$ in both KISO and IBUKI events is
considered to be the $1p$ state,
although several spins are possible.
Here, the isospin dependence of the ${\Xi}N$ interaction (Lane potential)
is proportional to $1/A$ and has a weak effect.
Assuming that the initial state is the same in both KISO and IBUKI events,
the weighted average
of the binding energy of $\Xi^-$ in the $1p$ state
is obtained to be $1.13 \pm 0.14$ MeV for the $\Xi^-$--$^{14}$N system.
This now gives the depth of the $\Xi^-$ potential for the first time.
On the other hand, in the case of the binding energy of $\Xi^-$ in
the $1p$ state being 3.87 $\pm$ 0.21 MeV in the KISO event,
the indicated width is too wide, despite the large contribution of
the Coulomb potential.
Thus, the present result is the first observation of
the Coulomb-assisted bound state for the $\Xi^-$--$^{14}$N system.
The probabilities of $\Xi^-$ hyperon capture from the $s$, $p$, and $d$ orbits
for $^{14}$N atom were estimated to be 0.00--0.07\%, 0.2--5.7\%, and
47.9--75.7\%, respectively \cite{cascade, koike}.
Therefore,
the observation of a $\Xi^-$ capture event in the $p$ orbit experimentally
indicates
the ${\Xi}N$--$\Lambda\Lambda$ coupling is weak, which agrees with
the recent study of the lattice QCD calculations \cite{lattice}.

In summary,
the J-PARC E07 experiment
observed a twin-$\Lambda$ hypernuclei event, named IBUKI.
The reaction process was clearly identified as
$\Xi^- + {}^{14}\text{N}\rightarrow{}^{10}_\Lambda\text{Be}
+ {}^{5}_\Lambda\text{He}$.
The binding energy of the $\Xi^- + {}^{14}$N system
was determined to be 1.27 $\pm$ 0.21 MeV by applying kinematic fitting.
By considering an excited state,
the energy level for $^{10}_\Lambda$Be was interpreted to be
the ground state ($1^-$) or the ($2^-$) spin doublet partner.
This is the first observation of twin-$\Lambda$ hypernuclei
in which the binding energy is precisely determined.
By considering the experimental data and the theoretical calculations,
the energy level of $\Xi^-$ is likely the Coulomb-assisted nuclear
$1p$ state for both the KISO and IBUKI events.
Assuming the same initial state for both events,
a binding energy of 1.13 $\pm$ 0.14 MeV was obtained as the weighted average.
Furthermore, the observation of a $\Xi^-$ capture event in the $p$ orbit
indicates that the ${\Xi}N$--$\Lambda\Lambda$ coupling is weak.


\begin{acknowledgments}
We acknowledge experimental supports from the staff members of
the J-PARC accelerator and the Hadron experimental facility.
We are grateful to the FUJIFILM Corporation for production of emulsion gel.
We thank the KEK computing research center for the computational resource.
We also thank National Institute of Informatics for SINET5.
We would like to express our gratitude to the staff of the Kamioka Observatory,
ICRR, for protecting our emulsion sheets in Kamioka mine against cosmic rays.
This work was supported by Japan Society for the Promotion of Science (JSPS)
KAKENHI Grants No. 23224006, No. JP16H02180, and No. JP20H00155,
and Ministry of Education, Culture, Sports, Science and Technology (MEXT)
KAKENHI Grants No. 15001001 (Priority Area), No. 24105002, No. JP18H05403, and
No. JP19H05417 (Innovative Area 2404), and National Research Foundation (NRF) of
Korea with Grants No. 2018R1A2B2007757.
S.B., J.P., and F.S. are supported
by Deutscher Akademischer Austauschdienst (DAAD) PPP Japan 2017 Grant No. 57345296
and by the European Union's Horizon 2020 research and innovation programme
under Grant Agreement No. 824093.
\end{acknowledgments}

\bibliography{ibuki-paper}


\end{document}